\documentclass[12pt,a4paper]{JHEP3}
\usepackage{graphics,amsmath}

\title{Symmetry of Anomalous Dimension Matrices for Colour Evolution of
  Hard Scattering Processes}
\author{Michael H. Seymour\\
School of Physics \& Astronomy, University of Manchester, U.K.; and\\
Theoretical Physics Group, CERN, CH-1211 Geneva 23, Switzerland.}

\abstract{In a recent paper, Dokshitzer and Marchesini rederived the
  anomalous dimension matrix for colour evolution of $\mathrm{gg \to
  gg}$ scattering, first derived by Kidonakis, Oderda and Sterman.  They
  noted a weird symmetry that it possesses under interchange of internal
  (colour group) and external (scattering angle) degrees of freedom and
  speculated that this may be related to an embedding into a context
  that correlates internal and external variables such as string theory.

  In this short note, I point out another symmetry possessed by all the
  colour evolution anomalous dimension matrices calculated to date.  It
  is more prosaic, but equally unexpected, and may also point to the
  fact that colour evolution might be understood in some deeper
  theoretical framework.  To my knowledge it has not been pointed out
  elsewhere, or anticipated by any of the authors calculating these
  matrices.  It is simply that, in a suitably chosen colour basis, they
  are complex symmetric matrices.}

\preprint{CERN-PH-TH/2005-157}

\begin{document}
\noindent
Following the pioneering work of Botts and Sterman\cite{Botts:1989kf},
it has long been known that all-orders resummation of virtual
corrections to scattering processes can be viewed as an evolution in the
colour space of the hard partons.  For all but the simplest scattering
processes, this colour space is non-trivial and the anomalous dimensions
for this evolution are non-commuting matrices, whose dimensionality
depends on the colour representations of the hard partons.  All the
massless $2\to2$ anomalous dimension matrices have been calculated in
\cite{Kidonakis:1998nf}.  They range from $2\times2$ matrices for
processes involving only quarks and antiquarks, to $9\times9$ for
$\mathrm{gg\to gg}$ (although 3 of these dimensions turn out to decouple
from physical amplitudes, and 1 is null for $\mathrm{SU}(N_c=3)$).  The
anomalous dimension matrices have also been calculated for massive
$\mathrm{Q\overline Q}$ production processes in \cite{Kidonakis:1997gm}.

More recently, Dokshitzer and Marchesini have rederived the most
complicated case of $\mathrm{gg\to gg}$
scattering\cite{Dokshitzer:2005ek} using a physically more transparent
method.  This allowed them to eliminate the unphysical decoupling
dimensions from the start, and to write the final result in a simpler
form.  In particular, this lays bare a weird symmetry that three of its
eigenvalues possess, and they speculated that this symmetry may point to
the possibility of a deeper theoretical understanding of colour
evolution, perhaps from string theory.

A deeper understanding of the colour evolution is badly needed if we are
to extend calculations to higher orders.  In general for an $n$-parton
scattering process (i.e.~$2\to(n\!-\!2)$), there are of order $n!$
colour states and the anomalous dimensions are extremely large matrices.
For example, for $\mathrm{gg\to ggg}$ there are 44 states for general
$\mathrm{SU}(N_c)$ and hence we must calculate and diagonalize a
$44\times44$ matrix.  Clearly some more fundamental organizing principle
is needed to make progress.

In this short note, I would like to point out another symmetry that all
the anomalous dimension matrices calculated to date possess.  To my
knowledge this symmetry was not anticipated by any of the other authors
working in the field, and was certainly not by me.  It is simply an
empirical observation on my part.

One is free to define the anomalous dimension matrix in any convenient
colour basis, and different calculations in the literature have used
different bases.  However, it is convenient to use an orthogonal basis
in which the lowest order soft matrix (in the nomenclature of
Refs.~\cite{Kidonakis:1998nf,Kidonakis:1997gm}, or the metric tensor of
the colour space in the nomenclature of Ref.~\cite{Dokshitzer:2005ek})
is diagonal, and all published results have done this.

As a concrete example, I use the result for $\mathrm{gg\to gg}$ using
the nomenclature of Kidonakis, Sterman et al, and the simplified form of
the result from Dokshitzer and Marchesini.  They use a set of
$s$-channel projectors as their colour basis, and therefore have a
lowest order colour matrix of
\begin{equation}
  \mathrm{S} = \left(\begin{array}{cccccc}
    K_{8_a}\\&K_{10}\\&&K_{1}\\&&&K_{8_s}\\&&&&K_{27}\\&&&&&K_{0}
  \end{array}\right),
\end{equation}
where $K_\alpha$ are the dimensionalities of the corresponding
representations,
\begin{equation}
\begin{split}
  K_1 &= 1, \qquad  K_{8_a}     = K_{8_s} = N_c^2-1, \qquad
  K_{10} = 2\cdot \frac{(N_c^2-1)(N_c^2-4)}{4}, \\
  K_{27} &= \frac{N_c^2(N_c-1)(N_c+3)}{4}, \qquad\qquad \quad
  K_{0}  \>=  \frac{N_c^2(N_c+1)(N_c-3)}{4},
\end{split}
\end{equation}
where the labels are the $\mathrm{SU}(3)$ values (and strictly speaking
10 should be $10+\overline{10}$).  They find an anomalous dimension
matrix of
\begin{equation}
 \Gamma = \left( \begin{array}{rrrrrr} \frac32 & 0 & \>\> -2b & \>\>
 -\frac{1}{2} b & -\frac{2}{N_c^2}b & -\frac{2}{N_c^2}b \\[2mm] 0 & 1 & 0
 & -b & \>\> -\frac{(N_c+1)(N_c-2)}{N_c^2}b & \>\>
 -\frac{(N_c-1)(N_c+2)}{N_c^2}b
 \\[2mm] -\frac{2}{N_c^2-1}b & 0 & 2 & 0 & 0 & 0 \\[2mm] -\frac12 b&
 -\frac{2}{N_c^2-4}b & 0 & \frac32 & 0 & 0 \\[2mm] -\frac{N_c+3}{2(N_c+1)}b
 & \>\> -\frac{N_c+3}{2(N_c+2)}b & 0 & 0 & \frac{N_c-1}{N_c} &0\\[2mm]
 -\frac{N_c-3}{2(N_c-1)}b & -\frac{N_c-3}{2(N_c-2)}b & 0 & 0 & 0 &
 \frac{N_c+1}{N_c} \end{array} \right),
\end{equation}
where
\begin{equation}
  b=\frac{T-U}{T+U},
  \qquad T=\ln\frac{s}{|t|}-i\pi,
  \qquad U=\ln\frac{s}{|u|}-i\pi.
\end{equation}

The symmetry I wish to point out is rather simple.  It is that in an
orthonormal basis, i.e.~one in which the lowest order soft matrix is the
identity matrix, the anomalous dimension matrix is symmetric.  Returning
to the concrete example of $\mathrm{gg\to gg}$, transforming to an
orthonormal basis, i.e.~normalizing the basis vectors, effectively
multiplies the $\alpha\beta$ element of $\Gamma$ by
$\sqrt{K_\beta/K_\alpha}$, yielding
\begin{equation}
 \Gamma =
 \left(
 \scalebox{0.68}{$\displaystyle
 \begin{array}{rrrrrr} \frac32 & 0 &
   \>\> -\frac{2}{\sqrt{N_c^2-1}}\,b & \>\>
 -\frac{1}{2} \,b & -\frac1{N_c}\sqrt{\frac{N_c+3}{N_c+1}}\,b &
 -\frac1{N_c}\sqrt{\frac{N_c-3}{N_c-1}}\,b \\[2mm]
 0 & 1 & 0
 & -\sqrt{\frac2{N_c^2-4}}\,b & \>\>
 -\sqrt{\frac{(N_c+3)(N_c+1)(N_c-2)}{2N_c^2(N_c+2)}}\,b & \>\>
 -\sqrt{\frac{(N_c-3)(N_c-1)(N_c+2)}{2N_c^2(N_c-2)}}\,b \\[2mm]
 -\frac{2}{\sqrt{N_c^2-1}}\,b & 0 & 2 & 0 & 0 & 0 \\[2mm]
 -\frac12 \,b& -\sqrt{\frac2{N_c^2-4}}\,b & 0 & \frac32 & 0 & 0 \\[2mm]
 -\frac1{N_c}\sqrt{\frac{N_c+3}{N_c+1}}\,b
 & \>\> -\sqrt{\frac{(N_c+3)(N_c+1)(N_c-2)}{2N_c^2(N_c+2)}}\,b & 0 & 0 & \frac{N_c-1}{N_c} &0\\[2mm]
 -\frac1{N_c}\sqrt{\frac{N_c-3}{N_c-1}}\,b &
 -\sqrt{\frac{(N_c-3)(N_c-1)(N_c+2)}{2N_c^2(N_c-2)}}\,b & 0 & 0 & 0 &
 \frac{N_c+1}{N_c} \end{array}
 $}
 \right),
\end{equation}
a manifestly symmetric matrix.

This is just one example, but I have found it to be true of all the
anomalous dimension matrices calculated to date.  It is true not only
for the anomalous dimension matrices of the virtual matrix elements, but
also those for energy flow observables, whether defined for a slice in
rapidity\cite{Oderda:1999kr}, a patch in rapidity and
azimuth\cite{Berger:2001ns}, or for a region of the event defined by the
$k_\perp$ algorithm\cite{Appleby:2003sj}.  In a forthcoming
paper\cite{Coulomb}, we have calculated an anomalous dimension matrix
for the colour evolution of a five-parton system, necessary to describe
a $2\to3$ process, as far as I know for the first time.  It also has
this symmetric structure in an orthonormal basis.

It is worth emphasizing that the anomalous dimension matrix is complex,
so the fact that it is symmetric is not equivalent to it being
Hermitian, a property that might have been slightly less surprising.  If
one wished to view the anomalous dimension matrix as a transition
amplitude for an effective theory in colour space, one might imagine
that time reversal symmetry of this theory would lead to a Hermitian
matrix, but the fact that it is symmetric rather than Hermitian does not
allow this interpretation.

I would like to close this short note by recalling that I have no
explanation for why the anomalous dimension matrices of colour evolution
should be symmetric.  This surprising result may turn out to be trivial,
or may point the way to a deeper understanding of colour evolution.


\begin{thebibliography}{9}

\bibitem{Botts:1989kf}
J.~Botts and G.~Sterman,
Nucl.\ Phys.\ B {\bf 325} (1989) 62.

\bibitem{Kidonakis:1998nf}
N.~Kidonakis, G.~Oderda and G.~Sterman,
Nucl.\ Phys.\ B {\bf 531} (1998) 365
[arXiv:hep-ph/9803241].

\bibitem{Kidonakis:1997gm}
N.~Kidonakis and G.~Sterman,
Nucl.\ Phys.\ B {\bf 505} (1997) 321
[arXiv:hep-ph/9705234].

\bibitem{Dokshitzer:2005ek}
Yu.~L.~Dokshitzer and G.~Marchesini,
arXiv:hep-ph/0508130.

\bibitem{Oderda:1999kr}
G.~Oderda,
Phys.\ Rev.\ D {\bf 61} (2000) 014004
[arXiv:hep-ph/9903240].

\bibitem{Berger:2001ns}
C.~F.~Berger, T.~Kucs and G.~Sterman,
Phys.\ Rev.\ D {\bf 65} (2002) 094031
[arXiv:hep-ph/0110004].

\bibitem{Appleby:2003sj}
R.~B.~Appleby and M.~H.~Seymour,
JHEP {\bf 0309} (2003) 056
[arXiv:hep-ph/0308086].

\bibitem{Coulomb}
J.~R.~Forshaw, A.~Kyrieleis and M.~H.~Seymour,
{\it On the R\^ole of Coulomb Phase Terms in Gaps Between Jets Cross
  Sections},
in preparation.

\end{thebibliography}
\end{document}